\documentclass[letterpaper,aps,twocolumn,superscriptaddress,tightenlines,
floatfix]{revtex4}
\usepackage{graphicx} 

\begin{document}

\title{Measuring the quantum efficiency of single radiating dipoles using a scanning mirror}

\author{B.~C.~Buchler}
\affiliation{Laboratory of Physical Chemistry,
Swiss Federal Institute of Technology (ETH), 8093 Zurich,
Switzerland.}
\author{T.~Kalkbrenner}
\altaffiliation[Present address:]{FOM-Institute for Atomic and Molecular Physics (AMOLF), 1098 SJ
Amsterdam, The Netherlands}
\affiliation{Laboratory of Physical
Chemistry, Swiss Federal Institute of Technology (ETH), 8093 Zurich,
Switzerland.}
\author{C.~Hettich}
\altaffiliation[Present address:]{Niels Bohr Institute, 2100 Copenhagen, Denmark}
\affiliation{Laboratory of Physical Chemistry,
Swiss Federal Institute of Technology (ETH), 8093 Zurich,
Switzerland.}
\author{V.~Sandoghdar}
\email{vahid.sandoghdar@ethz.ch}
\affiliation{Laboratory of Physical Chemistry, Swiss Federal Institute
of Technology (ETH), 8093 Zurich, Switzerland.} \pacs{}

\begin{abstract}
Using scanning probe techniques, we show the controlled manipulation
of the radiation from single dipoles.  In one experiment we study the
modification of the fluorescence lifetime of a single molecular dipole
in front of a movable silver mirror.  A second experiment demonstrates
the changing plasmon spectrum of a gold nanoparticle in front of a
dielectric mirror.  Comparison of our data with theoretical models
allows determination of the quantum efficiency of each radiating
dipole.

\end{abstract}
\date{\today} \maketitle

It is a well established matter that the radiation of an oscillating
electric dipole can be manipulated if it is placed in front of a
planar interface \cite{Barnes98}.  Experiments investigating this
system date back to Drexhage \cite{drexhage01} who looked at the
influence of a metallic mirror on the fluorescence lifetime of
ensembles of Eu$^{3+}$ ions.  By preparing a large number of samples,
each with a different spacing between the mirror and the emitter
layer, two major effects were observed.  Firstly, it was shown that
the decay rate ($\Gamma$)  oscillates at large distances due to the
retarded interaction of the dipoles with their own reflected fields.
Secondly, it was shown that $\Gamma$ is strongly modified very close
to the mirror due to the energy transfer to the
metal~\cite{drexhage01, Barnes98}.  Since that time, numerous works
have investigated the modification of spontaneous emission from
ensembles in thin dielectric layers \cite{Barnes98}.  Various key
parameters such as the dipole's orientation, its distance to the
interface and its quantum efficiency are, however, averaged in
ensemble measurements.

Due to challenges such as detection sensitivity, photostability and
position control, experiments with single emitters have been scarce.
Some researchers have nevertheless shown effects of the local
dielectric environment by adding an index matching fluid to eliminate
an interface \cite{Macklin_S1996, Brokmann_PRL2004} or by introducing
the subwavelength boundary of a sharp tip~\cite{Ambrose_S1994,
Trabesinger-02}.  In this work we study the fluorescence lifetime and
intensity of a single molecule at a well-defined orientation and
position, while moving an external silver mirror in its vicinity.  We
also examine the plasmon spectrum of a well-characterized single gold
nanoparticle at various locations in front of a dielectric mirror.
These experiments allow us to demonstrate, for the first time, both
the far-field modulation and the near-field modification of the total
decay rate ($\Gamma$) for individual dipoles.  Since the far-field
modulations are only due to changes in the radiative decay rate ($\Gamma_{r}$) we
can determine the quantum efficiency $\eta = \Gamma_{r}/\Gamma$ of
each dipole.

A theoretical description of dipole decay in multi-layer structures was first developed by Chance et al.~\cite{chance01} and has been expanded by many authors to cover numerous situations.  In particular, Sullivan and Hall~\cite{Sullivan_JOSAB1998} present an elegant plane wave solution that can easily be adapted to our system.
For a single dipole at angle $\theta$ with respect to the normal of the dielectric layers
we find \cite{Novotny1997_JOSAA}:
\begin{equation}
\Gamma = \Gamma_{0}[1-\eta_{0}+\eta_{0} (V \cos^2 \theta+
H\sin^2\theta)]\label{life}
\end{equation}
where $\eta_{0}$ and $\Gamma_{0}$ are the quantum efficiency and decay
rate, respectively, in bulk dielectric.  The functions $V$ and $H$ are, respectively,  the normalised decay rates of dipoles perpendicular and parallel to the layers. These functions depend on the position of the dipole, as well as the thickness and
refractive indices of the various layers.

\begin{figure}[t!]  \centering
\includegraphics[width=8.5 cm]{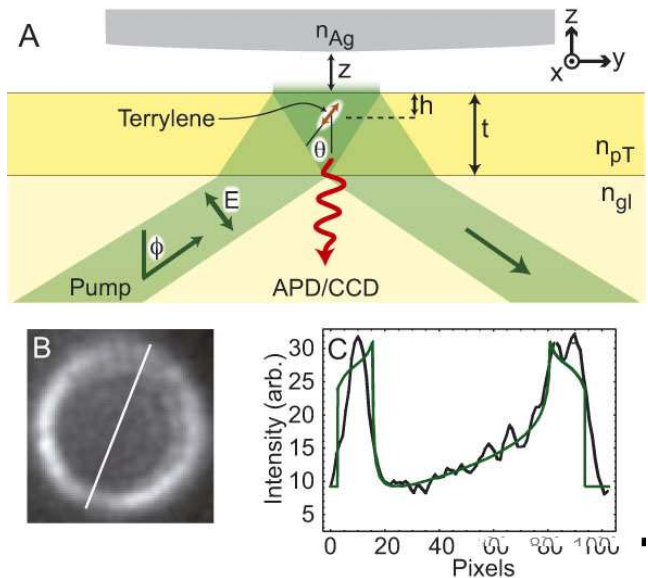}
\caption{A: Terrylene molecules embedded in para-terphenyl (pT) were
illuminated by p-polarized light that undergoes total internal
reflection at the pT-air interface.  A silver micro-mirror was positioned in
front of the sample with a three-dimensional piezo stage (not shown).
Fluorescence light is collected and sent to an APD or CCD camera.
Parameters used in the modelling are defined in this diagram.  B:~The
intensity distribution at the back focal plane for a single molecule
(with no mirror).  C:~A cross section indicated in B. A fit to this
data shows a dipole angle $\theta$ of 16$\pm$5 degrees.  }
\label{setup}
\end{figure}

The essential feature of our experimental system was the combination of an inverted microscope and a shear-force controlled scanning probe stage~\cite{karrai-00PRB}.  A schematic of the setup used for the single molecule measurements is shown in Fig.~\ref{setup}A. Samples of highly photostable terrylene molecules embedded in a thin para-terphenyl (pT) matrix were prepared following the method described in \cite{pfab01}. The thickness ($t$) of the pT layer in this work was 35$\pm5$nm as measured using shear-force microscopy.   The sample was illuminated through an oil immersion objective (NA=1.4) by 13~ps pulses of p-polarized 532~nm light at a
repetition rate of 5~MHz.  The laser light was focused in the back
focal plane of the objective to give a near-collimated beam at the pT
layer.  An offset of the beam at the back focal plane resulted in an
incidence angle $\phi$ and total internal reflection at the pT-air
interface.  The fluorescence of the terrylene ($\lambda\sim 580~nm$)
was collected confocally through the objective and sent to an
avalanche photodiode (APD) for lifetime measurements or to a CCD for
imaging.  Above the pT layer we mounted a micro-mirror in the shear-force stage.  The mirror was made by melting the end of a tapered optical fibre to give a ball of diameter 40 $\mathrm{\mu m}$, then coating this ball with 200 nm of silver.   Using the scanning stage, the mirror was lowered onto the pT layer, bringing it into
shear-force contact at a height $z_{0}$.  This point of closest
separation was in the range 5 to 20~nm.  The calibrated piezo used in
the z-axis of the system thus gave an accurate measurement of the
mirror-pT distance ($z$) relative to the zero defined by $z_{0}$.  The
lateral position of the mirror against a molecule was optimized by
maximizing the quenching of the fluorescence with the mirror at
$z_{0}$.

Figure~\ref{setup}B shows the image obtained when the fluorescence from a single molecule, measured without the mirror, is collected by the objective and sent directly to the  CCD with no further imaging optics.
 The highly directional (doughnut like) emission stems from
the fact that the terrylene dipole is oriented nearly perpendicular to
the substrate plane~\cite{pfab01}.  As has been shown very recently
\cite{Lieb_JOSAB2004}, by analyzing this image we can determine the
tilt angle $\theta$ of the dipole.  From the cross section in
Fig.~\ref{setup}C we find $\theta=16\pm 5 $ degrees.

Fluorescence lifetime ($\tau=1/\Gamma$) was measured via
time-correlated single photon counting.  Photon arrival times were histogrammed and
an exponential decay curve was fitted to find the excited state
lifetime.  To achieve the required accuracy, 1 to 2 s of data were required per measurement.  In bulk pT the lifetime of terrylene is
$\tau_{0}=1/\Gamma_{0}=4.1\pm0.1$~ns~\cite{harms01}.  In our system (without the micro-mirror) lifetimes were in the range 15 to 25 ns. Increased lifetime is expected for molecules embedded in thin dielectric films \cite{kreiter01}.  This effect can be described using Eq.~\ref{life} and, due to the dependencies of $V$ and $H$, is
highly sensitive to the dipole orientation and depth of the molecule
in the film.  A particular virtue of single emitter studies is their
ability to deal with such inhomogeneities.

\begin{figure}[b!]
\centering
\includegraphics[width=8.5 cm]{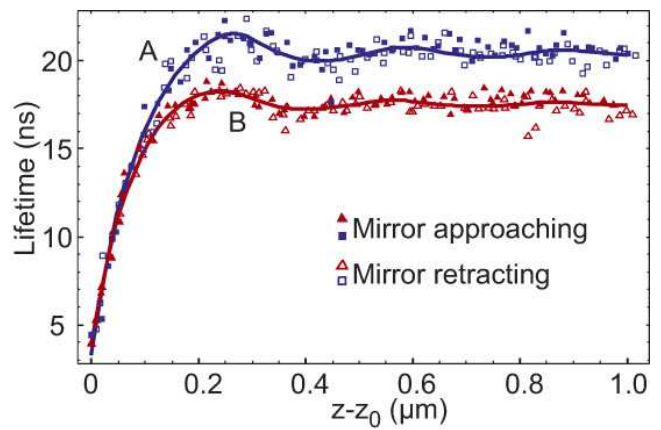}
\caption{Lifetime as a function of mirror position ($z$) for 2 molecules
A (squares) and B (triangles).  The solid curves display theoretical fits assuming $\eta_{0}=0.98$, $h=3~nm$ and $z_{0}=12~nm$ for A, $\eta_{0}=1$, $h=22~nm$ and
$z_{0}=11~nm$ for B and $\theta=16^\circ$ in both cases. } \label{lifetimes}
\end{figure}

We now turn to the lifetime measurements of single molecules as a function of the mirror position $z$.  During the course of the experiment, data was collected from around 15 different molecules.  Shown in Fig.~\ref{lifetimes} are two of these measurements.  The lifetime was monitored as the mirror moved towards and away from the molecule, thereby verifying the mechanical
stability of the system and lack of any drifts.  We note that more than $5\times10^7$ photons were detected per molecule, emphasizing the advantage of a sample with very high photostability~\cite{pfab01}.

The data in Fig.~\ref{lifetimes} clearly shows the near-field
shortening of the lifetime due to quenching by the metal as well as
far-field oscillations that can be understood as being due to the
retarded interaction of the dipole with its mirror
image~\cite{drexhage01,Barnes98}.  The solid curves in the figure show
fits to the data made using Eq.  \ref{life}.  The refractive indices
$n_{gl}$ and $n_{Ag}$ of glass and silver at 580~nm are taken to be
1.52 and 0.26+4$i$, respectively.  Para-terphenyl is  bi-axial with refractive indices $n_{z}=2.0$, $n_{y}=1.69$ and
$n_{x}=1.59$~\cite{pfab01,Sundararajan_ZC1936}. In the calculations we take the out-of-plane polarizations to have an index of $(n_{x}+n_{y})/2=1.64$. For polarizations parallel to the plane of incidence we assume an index of 1.85~\cite{footpt}.  Using these assumptions, the unknown parameters $\eta_{0}$, $h$ and $z_{0}$ can be approximated by fitting the model to the data.  For molecule A we find $0<h<13$~nm, $0.9<\eta_{0}< 1$ and $z_{0}=12\pm$2~nm whereas for B, $12<h<32$~nm, $0.9<\eta_{0}<1$ and $z_{0}=11\pm$5~nm. These values are consistent with what we know about our system. Molecules with longer lifetimes are expected to have smaller depth ($h$)
and the offset ($z_{0}$) agrees well with the typical shear-force
interaction range of $z_{0} < 20$nm.  The estimates of $\eta_{0}$ and
$h$ are limited by the uncertainty in the dipole orientation $\theta$,
a parameter that is integrated out in the theoretical treatments of
ensembles, but plays a central role in single emitter
studies~\cite{Brokmann_PRL2004}.

\begin{figure}[t!]
\centering
\includegraphics[width=8.5 cm]{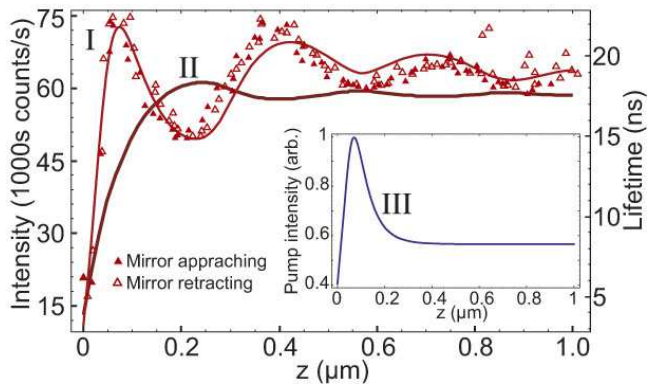}
\caption{I (left-hand axis): Symbols show measured intensity of molecule B
in Fig. ~\ref{lifetimes}.  The curve is the theoretical fit to the data with fitted parameters of $\phi=53\pm1$, $NA=1.00\pm0.01$ and other variables as in Fig~\ref{lifetimes}.  II (right-hand axis): The curve B of Fig.~\ref{lifetimes}.  III: The pump intensity at the position of the molecule with $\phi=53^\circ$.} \label{intensity}
\end{figure}

The fluorescence intensity is also measured using the APD.  For molecule B of Fig.~\ref{lifetimes} this is shown in Fig.~\ref{intensity}I along with the lifetime data (curve II).  For large $z$, there is a clear correlation between the intensity and lifetime.  The strong intensity modulation is due to our detection optics only seeing a part of the emission pattern while the angular distribution of the fluorescence is changing with the mirror position. The modification of emission patterns in the presence of planar boundaries was originally described by Drexhage~\cite{drexhage01} and more recently exploited in the design of semiconductor cavities~\cite{benisty:98a}.

The fluorescence intensity can be modelled by finding the emission
pattern and then integrating over the numerical aperture of the
collection optics \cite{Benisty_JOSAA1998,Enderlein_CP1999}.  As shown
by Fig.~\ref{intensity}III, we must also consider the change in pump
intensity experienced by the molecule when the mirror enters the
evanescent field above the pT surface.  As our measurements were made
well below the pump saturation level, we just multiply the integrated
emission pattern by the pump intensity.  The solid curve in
Fig.~\ref{intensity}I shows the theoretical fit to the intensity data,
displaying reasonable agreement.  We note that although the numerical
aperture of our objective was 1.4, the intensity modulation is
best reproduced by NA=1.  Numerous factors, such as misalignment of the detection optics or the birefringence of the pT film, could lead to a reduction of
the effective numerical aperture.  We believe that the main cause of
the discrepancy is that the integration of the emission pattern over
the numerical aperture assumes the entire dipole-mirror system is at
the focus of the objective.   This is not actually the case for
$z>500~nm$, equivalent to a Rayleigh range of the microscope
objective.  In this regime our high NA confocal detection system
begins to lose some of the photons that are first emitted upward and
then reflected by the mirror.

We will now consider the second experimental system, namely a gold
nanoparticle in front of a glass substrate.  A gold particle can
support plasmon resonances that stem from the collective oscillation
of electrons in the metal~\cite{Kreibig}.  The spectrum of the plasmon
resonance depends on the size and shape of the particle, as well as
the complex dielectric constants of the particle and surrounding
medium.  In general, the scattering and absorption properties of a
nanoparticle can be described by a multipole expansion.  For gold
particles smaller than about 100~nm, it suffices to consider only a
dipolar oscillation if one accounts for the effect of radiation
damping~\cite{Wokaun:82}. In an ellipsoidal particle, up to 3 dipolar oscillations can be excited yielding independent
plasmon resonances at 3 different frequencies~\cite{Kreibig}.  In a
recent publication we have demonstrated a tomographic approach to
identify these resonances and find the orientation of the
particle axes in the laboratory frame~\cite{kalkbrenner04}.
Furthermore, in~\cite{Kalkbrenner-01} we presented a recipe for
attaching a single nanoparticle to the extremity of a sharp optical
fiber.  Here we exploit these techniques to first characterize an
ellipsoidal nanoparticle (nominal diameter of 80~nm) attached to a
fiber tip and then monitor the width of one of its plasmon resonances
as we move the particle relative to a glass substrate.

\begin{figure}[b!]
\centering
\includegraphics[width=8.5 cm]{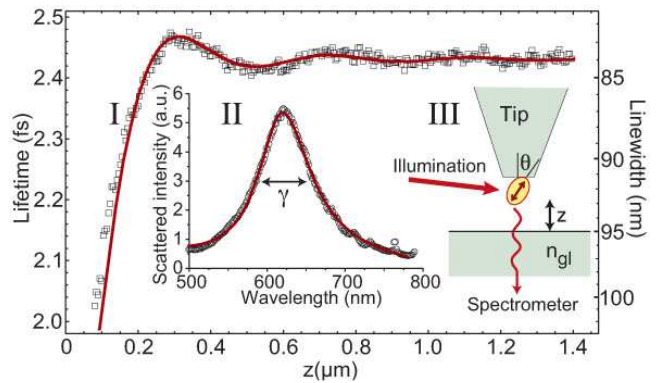}
\caption{I: The squares show the measured decay time and linewidth of the plasmon
oscillations.  The solid line shows a theoretical fit using Eq.~\ref{life}.  A stretch factor of 0.8 was fitted to the $z$ axis as the piezo in
this system was not calibrated.  II:~Circles show a plasmon spectrum
recorded from the nanoparticle at the tip.  The solid curve is a
fit according to a modified Mie theory.  III: ~A schematic of the
setup.} \label{goldtime}
\end{figure}

Details of the setup and the spectroscopy procedure are given in
\cite{kalkbrenner04}.  In short, the system was illuminated at grazing
incidence with white light from a xenon lamp (see
Fig.~\ref{goldtime}III).  The scattered light was collected by a
microscope objective (NA=0.85) through the glass substrate and sent to
a spectrometer.  By performing a tomographic measurement as described
in \cite{kalkbrenner04}, we determined the long axis of the particle
to be oriented at $\theta=10^\circ\pm2^\circ$.  We then rotated the
tip about its axis as well as the polarization of the illumination to
 excite only the long axis plasmon.
Figure~\ref{goldtime}II shows the measured plasmon spectrum and a fit obtained using the dipole term of Mie theory with a radiation damping correction~\cite{kalkbrenner04}.

A scanning piezo stage was used to control the distance ($z$) between
the gold particle and a microscope coverglass ($n_{gl}=1.52$).  The
point of closest approach ($z_{0}$) was found, as in the case of the
single molecules, using a shear force signal.  The
plasmon decay time was found according to $\tau=1/(2\pi\gamma)$~\cite{Soennichsen:02}, where $\gamma$  (in Hz)  is the full-width at half-maximum of the resonance.
A plot of the plasmon lifetime (and linewidth on the right-hand axis) as a function of $z$ is shown in Fig.~\ref{goldtime}I.
This result qualitatively resembles that of the molecular
dipole.  In particular for large $z$ we see slowly dying oscillations of the lifetime that eventually relax to a value of 2.43~fs.  As shown by the solid curve in Fig.~\ref{goldtime}I, the data is well reproduced by a theoretical fit according to Eq.~\ref{life}~\cite{footfibre}.  Fitted parameters in
this case were $z_{0}=9\pm6$~nm and $\eta_{0}=0.64\pm0.07$.  There are two complications in this data  that are unique to the gold plasmon measurements.  Firstly, the agreement between theory and experiment deteriorates for $z<200$~nm.  We believe that this is because a gold particle close to the glass surface can be no longer treated as a point dipole.  Data with $z<250$~nm is therefore excluded from the fit. Secondly, the linewidth measurement is affected by interference between the plasmon scattering and stray excitation light \cite{inprep} that causes a small but constant modulation of the linewidth at large $z$.  This artifact, with peak-to-peak amplitude 0.007~fs and period  equal to half the plasmon centre wavelength, has been subtracted from the data.

The so-called ``quantum efficiency'' for a classical antenna, such as a
gold particle, can again be defined as $\eta =
\frac{\Gamma_{r}}{\Gamma}$ whereby here $\Gamma_{r}$ and $\Gamma$
denote the scattering (radiation) and total plasmon relaxation rates,
respectively ~\cite{Heilweil:85,Soennichsen:02}.  Equivalently, this
quantity can be written as $\eta =
\frac{\sigma_{s}}{\sigma_{s}+\sigma_{a}}$ where $\sigma_{s}$ and
$\sigma_{a}$ denote the scattering and absorption cross sections of
the particle, respectively.  Using the outcome of the tomographic
measurements and the specified particle size, we calculated the values
of $\sigma_{s}$ and $\sigma_{a}$ according to Mie theory and obtained
$\eta_{0}=0.69\pm0.07$.  Considering the substantial uncertainty in the parameters that enter into the Mie calculation, the agreement between the calculated quantum efficiency and that extracted from the experiment is very good.

In summary, we have demonstrated the \emph{in situ} modification of
the decay rate for a single quantum emitter as well as a classical
nano-antenna.  In both cases the manipulation of a movable external
mirror at large distances modifies the dipole's radiative decay rate
but leaves its nonradiative dissipation channels unaffected.  This
enables us to extract the quantum efficiency of a single dipole, a
quantity that is nontrivial to determine even for
ensembles~\cite{Crosby1971_JPC}.  Our approach could be extended to
various emitters of current interest, ranging from dye molecules and
semiconductor nanoparticles to nanotubes.  Such measurements would allow the investigation of fluctuations due to variations in the structure of the emitter or its local environment.

We thank A.F.~Koenderink and S. K\"uhn for assistance and fruitful discussions.  This work was supported by the Swiss National Science Foundation.

\end{document}